\documentstyle[12pt]{article}
\begin{document}

\input{epsf}


\def\balpha{\mbox{\boldmath $\alpha$}}
\def\bbeta{\mbox{\boldmath $\beta$}}
\def\bgamma{\mbox{\boldmath $\gamma$}}
\def\bomega{\mbox{\boldmath $\omega$}}

\newcommand{\nd}[1]{/\hspace{-0.5em} #1}
\begin{titlepage} 
\bigskip
\hskip 3.7in\vbox{\baselineskip12pt
\hbox{POLFIS-TH}\hbox{SWAT-168}\hbox{hep-th/9710098}}
\bigskip\bigskip\bigskip\bigskip

\centerline{\large \bf Instantons, Three Dimensional Gauge Theories,}  
\centerline{\large \bf and Monopole Moduli Spaces} 
\bigskip\bigskip
\bigskip\bigskip

\centerline{Christophe Fraser$^{1,2}$ and David Tong$^{2}$} 
\bigskip\bigskip

\centerline{$^{1}$Dipartimento di Fisica, Politecnico di Torino,}
\centerline{Corso Duca degli Abruzzi 24, 10129 Torino, Italy}

\vspace{.3in}

\centerline{$^{2}$Department of Physics, University of Wales Swansea }
\centerline{Singleton Park, Swansea, SA2 8PP, UK}
\centerline{pycf , pydt@swan.ac.uk}
\bigskip\bigskip

\begin{abstract}
\baselineskip=16pt
We calculate instanton corrections to three dimensional gauge theories 
with $N=4$ and $N=8$ supersymmetry and $SU(n)$ gauge groups. 
The $N=4$ results give new information about the moduli space of $n$ 
BPS $SU(2)$ monopoles, including the leading order non-pairwise 
interaction terms. A few comments are made on the relationship of 
the $N=8$ results to membrane scattering in matrix theory. 
\end{abstract}
\end{titlepage}

\section{Introduction}

In the past year a remarkable relationship between three 
dimensional gauge theories and monopole moduli spaces has been uncovered. 
Following the work of Seiberg 
and Witten \cite{sw}, Chalmers and Hanany \cite{ch} were the first to 
conjecture that the moduli space of $n$ BPS $SU(2)$ monopoles is equivalent 
to the vacuum moduli space of $SU(n)$ gauge theory in three dimensions 
with $N=4$ supersymmetry. This proposal found its natural setting in 
the work of Hanany and Witten \cite{hw}, where configurations of 5 branes 
and 3 branes in IIB string theory lead directly to the result.

The $SU(2)$ theory has subsequently been subjected to a first principles 
instanton calculation \cite{dkmtv}. In this case the vacuum moduli space 
is severely restricted by the (super)symmetries and perturbative sector of 
the theory, allowing for just a one parameter family of metrics. A one 
instanton calculation is sufficient to fix this parameter and the 
resulting metric is indeed that of the two monopole moduli space, 
known as the Atiyah Hitchin metric \cite{ah}.

In the following section we consider $N=4$ $SU(n)$ gauge theory 
in three dimensions. The corresponding $n$ monopole $SU(2)$ moduli 
space is known only for well-seperated monopoles \cite{gm}. We 
calculate instanton corrections in the three dimensional theory which 
correspond to the first exponential corrections to this metric.
In three dimensions the relevant instantons are BPS monopole configurations
\footnote{To avoid confusion we will refer to these configurations as 
``instantons'' with the term ``monopole'' reserved for the vacuum moduli 
space.}. We review such configurations in higher rank gauge groups 
with a Higgs field transforming under a global R-symmetry. 
In the presence of extra Higgs fields, the zero modes of 
instantons are fewer than the single Higgs 
results \cite{tim,wein} in a 
manner crucial for the interpretation of $n$-particle scattering. 
The non-zero modes around the background of the instanton are treated 
in the Gaussian approximation and, as in the $SU(2)$ case \cite{dkmtv}, 
there is a non-cancellation of bosonic and fermionic modes. However, 
unlike the situation for $SU(2)$ instantons, there 
exist {\em curves of marginal stability} (CMS) within the weak coupling 
regime of the moduli space of vacua upon which certain non-zero modes 
become zero-modes i.e. the instanton 
moduli space is enlarged. For these modes the Gaussian approximation is 
not sufficient and we treat them exactly using the method of 
constrained instantons \cite{aff}. 
A potential is introduced on the enlarged instanton moduli 
space reflecting the fact that these configurations are not in general 
solutions to the full equations of motion. We find the potential is 
generated by the norm of the $U(1)$ killing vectors of the instanton moduli 
space. 

In section 3, we translate these results into the language of the moduli 
space of $n$ monopoles of $SU(2)$ and find the leading order 
exponentially surpressed 
 corrections to the metric of Gibbons and Manton \cite{gm}. 
The non-cancellation of the instanton background fluctuations leads to a 
structure for the metric corrections corresponding to non-pairwise 
interactions between monopoles. These corrections become singular 
in the limit of co-linear monopoles due to the extra zero modes 
appearing on the CMS. These singularities are resolved by the 
constrained instanton approach and we find the expected behaviour 
in the limit of co-linearity. 

Further applications of three dimensional 
instantons have arisen in the context of Matrix theory \cite{bfss}. 
Polchinski and Pouliot \cite{pp} related the dynamics of 
two membranes scattering with momentum transfer in the longitudinal 
direction to instantons in three dimensional $SU(2)$ gauge theory, this time 
with $N=8$ supersymmetries. The $k$-instanton corresponds to $k$ units 
of transferred momentum. A one instanton calculation performed 
in \cite{pp} was found to be in agreement with the equivalent eleven 
dimensional supergravity calculation. Dorey, Khoze and Mattis \cite{dkm} 
later performed the all-instanton calculation, retaining agreement 
with supergravity. The $k$-instanton contribution is proportional to the 
Euler character of the $k$-monopole moduli space (up to certain boundary 
terms which are proposed to vanish). In section 4, we generalise 
this result to $SU(n)$ gauge groups. The extra supersymmetry 
means the background fluctuations now cancel between bosons and fermions, 
ensuring the corresponding membrane scattering acts in a pairwise manner.

\section{Three Dimensional Instantons} 

$N=4$ supersymmetric gauge theory in three dimensions is best viewed as 
the dimensional 
reduction of the six dimensional ${\cal N}=1$ theory. The bosonic sector 
contains the three dimensional gauge field, $A_\mu$, with field strength, 
$F_{\mu\nu}$, and three scalars, 
$\phi^i$, $i=1,2,3$. The scalars transform as a vector under a global 
$SO(3)$, the remanant of the six dimensional Lorentz  group. Following 
\cite{sw} we denote the double cover of this group as $SU(2)_N$. 

The Weyl fermion of six dimensions decomposes as four two-component 
Majorana fermions in three dimensions, $\chi^m_\alpha$, $m=1,..,4 ; 
\alpha=1,2$. There exists a second R-symmetry, denoted $SU(2)_R$, under 
which the scalars are singlets. The fermions transform under both global 
symmetry groups, as the ${\bf 4}$ of  $Spin(4)\simeq SU(2)_{N}\times SU(2)
_R$. All fields transform in the adjoint of the gauge group. 

As is usual in theories with extended supersymmetry, the scalar potential, 
$V(\phi)=\frac{1}{2}\sum_{i,j}[\phi^{i},\phi^{j}]^2$, has flat directions. 
The vacuum expectation value of the scalars are taken to live in ${\bf H}$, 
the $n-1$ dimensional Cartan subalgebra (CSA) of $SU(n)$\footnote{we use 
bold type to denote vectors in the root space and a superscript $i$ 
for 3-vectors transforming under $SO(3)_N$.}.
\begin{equation}\label{vev}
\langle\phi^{i}\rangle ={\bf v}^{i}\cdot{\bf H}\ \ \ \ \ ;\ \ \ \ \ i=1,2,3
\end{equation}
For maximal symmetry breaking, $SU(n)\rightarrow U(1)^{n-1}$, we require 
$\|{\bf v}^i\cdot{\balpha}\|\neq 0$ , for all roots $\balpha$ where 
$\|$ denotes the norm of the $SO(3)_N$ vector. 
This is assumed for the remainder of the paper.  

Unlike the situation with a single Higgs field, for a generic vacua, 
the vev's (\ref{vev}) do 
not pick out a unique set of simple roots, an observation at 
the heart of the zero mode structure for instantons in these 
theories. Although there is no unique choice, positive roots  
$\balpha^A,\, A=1,..,\frac{1}{2}n(n-1)$ may always be defined by choosing 
a suitable constant 3-vector, $\rho^i$, and requiring 
$\rho^{i}{\bf v}^{i}\cdot{\balpha}^{A}\geq 0$. We normalise the roots 
as $\balpha^{A}\cdot\balpha^{A}=1$ 
(no sum over $A$). Decomposing the fields into the Cartan-Weyl basis, 
those living along 
the step operators $E_{\pm A}$ pick up masses 
$M_{A} = \|{\bf v}^i\cdot\balpha^A\|$ by the adjoint Higgs mechanism. 
The fields living in the Cartan subalgebra remain massless. 
The choice of positive roots defines 
a set of simple roots, $\bbeta^a$, $a=1,..,n-1$, which we choose to 
define a (non-orthogonal) basis for the massless gauge fields.
\begin{equation}\label{csafields}
A_\mu^a = {\rm Tr}(A_{\mu}\bbeta^{a}\cdot{\bf H})\ \ \ \ \ ;\ \ \ \ \ 
a=1,...,n-1
\end{equation}
with similar definitions for the supersymmetric partners. 

Concerning ourselves just with the massless fields the classical 
approximation to the Euclidean 
low-energy Lagrangian is a free abelian theory, with bosonic sector
\begin{equation}\label{claslow}
S_B = \frac{2\pi}{e^2}\int d^{3}x (K^{-1})_{ab}\left(\frac{1}{4}
F^a_{\mu\nu} F^b_{\mu\nu} + \frac{1}{2}\partial_\mu \phi^{ia}\partial_\mu 
\phi^{ib} \right)
\end{equation}
where the inverse Cartan matrix, $K^{-1}$, makes an appearance as the 
metric of the classical sigma model. 

In the maximally broken abelian theory, a surface term is included to count 
the winding of the gauge field at infinity. Defining $n-1$ winding numbers,  
\begin{equation}\label{winding}
n_a=\frac{1}{8\pi}(K^{-1})_{ab}\int d^3x \epsilon_{\mu\nu\rho}\partial_\mu 
F^b_{\nu\rho} \ \ \in Z,
\end{equation}
the surface term is given by $S_S=in_a\sigma^a$. The parameters, 
$\sigma^a$, can 
be thought of as Lagrange multipliers for the $U(1)$ Bianchi identities 
and, as is clear from (\ref{winding}), they range from 0 to $2\pi$. 
Promoting each $\sigma^a$ to a dynamical field, we integrate out the field 
strengths in favour of these periodic scalars to obtain the dual description 
of the classical low energy effective action with $4(n-1)$ massless scalars 
and $4(n-1)$ massless Majorana fermions,
\begin{equation}\label{claslow2}
S=\frac{2\pi}{e^2}\int d^3x (K^{-1})_{ab}\left(\frac{1}{2}\partial_\mu\phi
^{ia}\partial_\mu\phi^{ib}+\frac{e^4}{\pi^2(8\pi)^2}\frac{1}{2}\partial_\mu
\sigma^a\partial_\mu\sigma^b+\frac{i}{2}\chi^{am}\gamma_\mu\partial_\mu\chi^
{bm}\right)
\end{equation}
where we take the three-dimensional gamma matrices, $\gamma_\mu$, to be the 
Pauli matrices, $(\sigma^3,-\sigma^1,\sigma^2)$.

Let us re-examine the symmetries of the low energy theory. The vevs 
generically spontaneously break the $SU(2)_N$ symmetry completely 
(for $SU(2)$ gauge group, there remains an unbroken $U(1)_N$). 
The low energy action, (\ref{claslow2}), has $n-1$ new abelian symmetries, 
$\sigma^a\rightarrow\sigma^a+c^a$ for any constants $c^a$. 
Because of the additive nature of this transformation, these too 
are spontaneously broken. At the classical level, the vacuum moduli space 
is ${({\bf R^3}\times{\bf S^1})^{n-1}/S_{n-1}}$ where $S_{n-1}$ is 
the Weyl group of $SU(n)$. This moduli space inherits the metric from the 
low-energy sigma model; classically, the inverse Cartan matrix acting on 
the $n-1$ copies of ${\bf R^3}\times{\bf S^1}$. 

The $4(n-1)$ massless scalars (and fermions) remain massless in the full 
quantum theory \cite{sw}. The Wilsonian low-enery 
effective action, obtained by integrating out all massive modes, replaces 
$\delta_{ij}\times K^{-1}_{ab}$ with the quantum corrected  metric 
$g_{ai\, bj}$, now depending on the vevs of 
$\phi$'s and $\sigma$'s, with $i,j=1,2,3,\sigma$.  
Four supersymmetries force $g_{ai\,bj}$ to be hyperkahler, 
while a non-anomalous $SO(3)_N$ global symmetry requires $g_{ai\,bj}$ 
to admit an $SO(3)$ isometry. It is proposed that $g_{ai\,bj}$ 
is the metric of the moduli space of $n$ BPS monopoles 
with $SU(2)$ gauge group \cite{sw,ch,hw}. This metric is known to be 
complete implying the singulaties of the classical vacuum moduli 
space arising as $M_A\rightarrow 0$ are resolved by strong coupling 
quantum effects. 

Perturbatively, the $U(1)$ symmetries shifting the $\sigma$'s are respected 
and corrections to the metric must contain $n-1$ abelian isometries. 
Chalmers and Hanany \cite{ch} have confirmed the perturbative corrections to 
$g_{ab}$ do indeed reproduce the asymptotic form of the $n$ monopole 
moduli space discovered by Gibbons and Manton \cite{gm}. In the monopole 
picture, the $U(1)$ isometries correspond to the conservation of electric 
charge of each individual dyon. Non-perturbatively, these $U(1)$ 
symmetries of the field theory are violated by instantons which in the 
monopole picture leads to charge exchange between dyons as their cores 
overlap. We now examine these instantons in more detail. 

\subsection*{Instanton Zero Modes}

For $SU(n)$ we have different species of three dimensional instanton 
labelled by their winding number (\ref{winding}), which we take to  
define a charge vector in the root lattice, ${\bf g}=n_a\bbeta^a$. 
The instantons of interest satisfy the Bogomol'nyi equation \cite{tim,ct}, 
\begin{equation}\label{bog}
{\cal D}_{\mu}\phi^{i}=\lambda^{i}_{\bf g}
B_{\mu}\ \ \ \ ;\ \ \ \ [\phi^i,\phi^j]=0
\end{equation}
where $B_{\mu}=\frac{1}{2}\epsilon_{\mu\nu\rho}F^{\mu\nu}$ and 
$\lambda^i_{\bf g}$ is given by
\begin{equation}\label{lambda}
\lambda^{i}_{\bf g}
=\frac{{\bf v}^{i}\cdot{\bf g}}{\|{\bf v}^{i}\cdot{\bf g}\|}
\end{equation}
Solutions of (\ref{bog}) have the property that they are annihilated by half 
the supersymmetries. The action of such a solution saturates the Bogomol'nyi 
bound and is given by 
\begin{equation}\label{action}
S_{\bf g}=\frac{8\pi^{2}}{e^{2}}\lambda^i_{\bf g}
{\bf v}^i\cdot{\bf g}+in_{a}\sigma^{a}
\end{equation}
The $in_{a}\sigma^{a}$ term was first introduced by Polyakov \cite{poly} to 
incorporate the long range effects of instantons in the dilute gas 
approximation. In the present context it appears through the 
surface term of the action, $S_S$.

A class of explicit solutions can be constructed by embedding charge $k$ 
$SU(2)$ instantons in the $SU(2)$ subgroup associated with $\balpha^A$. 
\begin{eqnarray}\label{subgroup}
t^1 & = & \frac{1}{\sqrt{2}}(E_A+E_{-A}) \nonumber \\
t^2 & = & \frac{1}{\sqrt{2}i}(E_A-E_{-A}) \\
t^3 & = & \balpha^A \cdot{\bf H} \nonumber
\end{eqnarray} 
with a ${\bf g}=k\balpha^A$ instanton solution obtained by,
\begin{eqnarray}\label{inst}
\phi^i & = & \lambda^i\phi^m (v)t^m + ({\bf v^i}-({\bf v^i}\cdot
\balpha^A)\balpha^A)\cdot{\bf H} \nonumber \\
A_\mu & = & A_\mu^m (v)t^m
\end{eqnarray}
where $\phi^m (v)$ and $A_\mu^m (v)$ are the solution for BPS monopoles in 
$SU(2)$ with a single Higgs field of expectation value
$v=\| {\bf v}^i\cdot{\bf \balpha^A}\|$  \footnote{For a detailed 
review of BPS monopoles as 3d instantons see Appendix C of \cite{dkmtv}.}. 
In fact, we will see that the extra Higgs fields ensure that for 
generic values of the expectation values all instantons are of this form. 
Below we apply the Callias index theorem and will 
infer that the only solutions of (\ref{bog}) have charge vector 
${\bf g}\propto\balpha^A$ for some root 
$\balpha^A$. This is in contrast to the situation with a single Higgs 
field where instantons exist for all charge vectors, ${\bf g}=
\sum_am_a\bbeta^a$, 
with $4\sum_am_a$ zero modes \cite{wein}. That this is no longer the situation 
here is known from an analysis of monopoles in four-dimensions \cite{tim,ct}, 
and can be anticipated from (\ref{vev}); the three vevs do not pick out a 
unique set of simple roots, $\bbeta^a$.

Rather than counting directly the number of bosonic zero modes, we determine 
the number of fermionic zero modes in the background of each instanton. 
The unbroken supersymmetry then pairs fermionic and bosonic zero modes. The 
Dirac equation reads,
\begin{equation}\label{dirac}
\Delta_{mn}\,\chi^n = (i\gamma_\mu{\cal D}_\mu\delta_{mn}-\eta^i_{mn}\phi^i)
\chi^n
\end{equation}
where $\eta^i$ are the self-dual t'Hooft matrices and the 
covariant derivative is 
referred to the background field of the instanton. 

Introducing the projection 
operators ${P_\pm =\frac{1}{2}(1\pm i\lambda^i_{\bf g}\eta^i)}$ acting on the 
$Spin(4)$ vector space, we take products with the adjoint operator, 
\begin{eqnarray}    
\Delta\Delta^\dagger =-{\cal D}_\mu{\cal D}_\mu-2\gamma_\mu B_\mu P_+ +\phi^i
\phi^i \nonumber \\ \Delta^\dagger\Delta =-{\cal D}_\mu{\cal D}_
\mu-2\gamma_\mu B_\mu P_- +\phi^i\phi^i 
\end{eqnarray}
Observing that $\Delta^\dagger\Delta P_+$ is positive definite, all zero 
modes of $\Delta$ must lie in the eigenspace of $P_-$ where $\Delta\Delta
^\dagger$ is itself positive definite. Let ${\rm Tr}_-$ be the trace function 
restricted to this space, and following Weinberg \cite{wein} define
\begin{equation}\label{Imu}
{\cal I}(\mu^2)={\rm Tr}_-\left(\frac{\mu^2}{\Delta^\dagger\Delta +
\mu^2}\right)-{\rm Tr}_-\left(\frac{\mu^2}{\Delta\Delta^\dagger +\mu^2}\right)
\end{equation}
The number fermionic zero modes is given by the limit $\mu^2\rightarrow 0$ 
of $2{\cal I}(\mu^2)$. We rewrite 
$\phi^i\phi^i=\Phi\Phi+\hat{\phi}^i\hat{\phi}^i$ where,
\begin{equation}\label{phihat} 
\Phi = \lambda^i_{\bf g}\phi^i\ \ \ \ ; \ \ \ \ \hat{\phi}^i = 
(\delta^{ij}-\lambda^i_{\bf g}\lambda^j_{\bf g})\phi^j
\end{equation}
Note that the three $\hat{\phi}^i$ have only two independant degrees of 
freedom. With the exception of the extra Higgs fields, 
$\hat{\phi}^i\hat{\phi}^i$, equation (\ref{Imu}) is the same as 
Weinberg's function \cite{wein}. In 
Appendix A we repeat Weinberg's calculation with this term and find 
\begin{equation}\label{337}
{\cal I}(\mu^2)=2\sum_{A}\frac{\mu^2\lambda^i_{\bf g}({\bf v}^i\cdot\balpha^A)
({\bf g}\cdot\balpha^A)}{(\|\hat{\bf v}^i\cdot\balpha^A\|^2 + \mu^2)
(\|{\bf v}^i\cdot\balpha^A\|^2 + \mu^2)^\frac{1}{2}}
\end{equation}
We now see the consequences of the extra Higgs fields for the zero mode 
structure. For most charges, ${\bf g}$, the factor of $\mu^2$ in the 
numerator of (\ref{337}) means there are no zero modes at all. For 
a non-zero ${\cal I}(M^2\rightarrow 0)$, we require,
\begin{equation}\label{onlyroot}
\|\hat{\bf v}^i\cdot\balpha^A\|^2=
({\bf v^i}\cdot\balpha^A)({\bf v}^i\cdot\balpha^A)-
\lambda^i_{\bf g}\lambda^j_{\bf g}
({\bf v}^i\cdot\balpha^A)({\bf v}^j\cdot\balpha^A)=0
\end{equation}
for some root $\balpha^A$ (no sum over $A$). For the case of generic vev, 
the only solution is 
${\bf g}$ aligned with $\balpha^A$, say ${\bf g}=k\balpha^A$. 
In this case we have 
\begin{equation}\label{zmodes}
\lim_{\mu^2\rightarrow 0}\ {\cal I}(\mu^2)=2k\frac{\lambda^i_{\bf g}
{\bf v}^i\cdot\balpha^A}{\|{\bf v}^i\cdot\balpha^A\|} = 2k
\end{equation}
For ${\bf g}$ not proportional to a root, equations (\ref{bog}) have no 
zero modes and so at most only isolated solutions. Yet a soliton 
must have translational zero modes and we infer that, for generic 
expectation values, the Bogomol'nyi equation has no solutions in these 
sectors.

In certain vacua known as curves of marginal stability (CMS)
\footnote{The name derives from studies of spectra in four dimensional 
${\cal N}=2$ theories where certain solitonic states are at threshold for 
decay. For gauge groups of rank $r\geq 3$ the CMS extend into the weak 
coupling regime.}, where 
$\lambda^i_A=\lambda^i_B$ for some $A\neq B$, 
equation (\ref{onlyroot}) has more solutions and the moduli space of 
the corresponding instanton enlarges. In the special case of 
$\lambda^i_A=\lambda^i_B$ for all roots 
$\balpha^A$ and $\balpha^B$, equation (\ref{337}) 
reduces to Weinberg's expression \cite{wein}. 

Thus the extra Higgs have two effects; enforcing a democracy amongst roots 
and removing any off-root solution to the Bogomol'nyi equations.  The former 
is already well known from studies of BPS spectra in four-dimensional theories 
with both ${\cal N}=2$ \cite{tim} and ${\cal N}=4$ \cite{ct} 
supersymmetries. The $\hat{\phi}^i\hat{\phi}^i$ terms punish any 
deviation from the $SU(2)$ subgroup (\ref{subgroup}), restricting the 
solution to be of the form (\ref{inst}). In \cite{ct} this is used to 
explain how, for generic vev, S-duality of the BPS spectrum of ${\cal N}=4$ 
theories with arbitrary simple gauge group is reduced to the equivalent 
problem in $SU(2)$, at least for charges proportional to roots. That 
no BPS solitons exist for charges not aligned with a root 
completes this argument.
\footnote{The above argument does not forbid ``middle multiplets'' 
with electric charge not parallel to magnetic charge. The zero modes 
for soltions with electric charge require a more delicate handling of 
the index theorem. We thank T. Hollowood for explaining this.}

For the three dimensional case in hand, this null result for the 
zero modes of instantons with charge not aligned with a root means 
the action of any instanton in these sectors is raised above the 
Bogomol'nyi bound and thus the configurations break all of the 
supersymmetries. Acting with these broken supersymmetries gives rise to too 
many fermionic zero modes to contribute to the low energy Wilsonian 
effective action with two derivatives and four fermions. Similarly, on the 
CMS, instantons in these sectors must have at least eight fermionic zero 
modes, again to many to contribute. Thus we restrict 
our attention to charges aligned with roots, ${\bf g}=k\balpha^A$, and 
denote $\lambda^i_{\bf g}=\lambda^i_A$.

Every such instanton has at least four bosonic modes, three corresponding 
to translations in space and time and one global $U(1)$ gauge transformation. 
The contribution of these modes to the bosonic measure is 
\begin{equation}\label{bmeasure}
\int d\mu_B =\int\frac{d^3 x}{(2\pi )^{\frac{3}{2}}}
({\cal J}_X)^{\frac{3}{2}}\int_0^{\frac{2\pi}{k}}
\frac{d\theta}{(2\pi)^{\frac{1}{2}}}({\cal J}_{\theta})^{\frac{1}{2}}
\end{equation}
where ${\cal J}_X$ and ${\cal J}_\theta$ are the Jacobians resulting from 
the change of variables from fields to zero modes. These are calculated in 
appendix C of \cite{dkmtv} (see also \cite{dkm}) for gauge group $SU(2)$. 
In the present  case the instanton is restricted to an $SU(2)$ 
subgroup and the calculation of \cite{dkmtv} generalises 
trivially. For the instanton with ${\bf g}=\balpha^A$, ${\cal J}_X = 
8\pi^2M_A/e^2$ and ${\cal J}_{\theta} = 8\pi^2/M_Ae^2$. 

Similarly, each configuration has at least 
two fermionic collective coordinates 
corresponding to broken supersymmetry generators. 
Supersymmetry transformations on the fermions with parameters 
$\xi^m_\alpha$ yield, 
\begin{equation}\label{fzero}
\chi^m=-i\gamma_\mu B_\mu (P_-)_{mn}\xi^n
\end{equation}
For instantons, the required broken supersymmetries have $\xi^m$ 
living in the eigenspace of the projection operator $P_-$. 
If $\xi^1$ and $\xi^2$ are the two eigenvectors of $P_-$, the contribution 
of these modes to the fermionic measure is
\begin{equation}\label{fmeasure}
\int d\mu_F = \int d^2\xi^1d^2\xi^2 ({\cal J}_\xi )^{-2}
\end{equation}
The fermionic Jacobians, ${\cal J}_\xi = 16\pi^2M_A/e^2$, are 
also calculated in appendix C of \cite{dkmtv}. 

The Grassmann integrations of (\ref{fmeasure}) are saturated by the 
insertion of four fermi fields in the path integral. If the instanton 
solution is to contribute to the low energy effective action at 
the two derivative and four fermi level, any 
further fermionic zero modes must be lifted. In section 4, we will 
discuss the case of an adjoint massless matter multiplet (N=8 supersymmetry) 
where such lifting does indeed occur \cite{dkm}. In the N=4 theory with 
no matter multiplets there is no mechanism for lifting extra 
fermionic zero modes and the only instantons that contribute must 
have the zero modes of equations (\ref{bmeasure}) and (\ref{fmeasure}) 
and no others. Thus we restrict our attention yet again 
to ${\bf g}=\balpha^A$ for each root ${\balpha^A}$. Further contributions 
come from perturbative (two-loop) corrections about the background 
of these solutions and various numbers of instanton-anti-instanton pairs.

\subsection*{Instanton Non-Zero Modes}

Before integrating over zero modes, we must first deal with the 
non zero fluctuations around the background of the instanton. 
Expanding about the configurations to quadratic order, the 
Gaussian integrations yield determinants of the quadratic fluctuation 
operators. In \cite{dkmtv} these were found to be non-trivial for the 
case of $SU(2)$ and for higher rank gauge groups we find even more 
structure. Choosing the background gauge, ${\cal D}_\mu\delta A_\mu -
i[\phi^i,\delta\phi^i]=0$, we find the contribution from the ghost fields 
cancels the fluctuations around $\hat{\phi}^i$. Supersymmetry ensures 
that the remaining bosonic and fermionic fluctuations are related and we find
\begin{equation}\label{dets}
R=\left(\frac{{\rm det}(-{\cal D}_\mu{\cal D}_\mu + \phi^i\phi^i)}
{{\rm det'}(-{\cal D}_\mu{\cal D}_\mu -2\gamma_\mu B_\mu + \phi^i\phi^i)}
\right)^{\frac{1}{2}} = \left(\frac{{\rm det}(\Delta\Delta^\dagger P_- +\Delta
^\dagger\Delta P_+)}{{\rm det'}(\Delta\Delta^\dagger P_+ +\Delta
^\dagger\Delta P_-)}\right)^{\frac{1}{2}}
\end{equation}
where the operator in the denominator has zero eigenvalues and ${\rm det'}$ 
denotes the removal of these from the determinant. Although supersymmetry 
insists the non-zero eigenvalues of the two operators in (\ref{dets}) are 
equal, the densities of these values are not. This was first 
noticed by Kaul \cite{k} in the context of mass renormalisation 
of monopoles. As explained in \cite{dkmtv}, the calculation of this 
ratio is essentially equivalent to the index calculation of the 
appendix. More precisely, for the instanton ${\bf g}=\balpha^A$, we 
have
\begin{eqnarray}\label{R}
R & = & \lim_{\mu\rightarrow 0}\left[\mu^{2}\exp\left(
\int^\infty_{\mu}\frac{d\nu}{\nu}{\cal I}(\nu)\right)\right]^{\frac{1}{2}} 
\nonumber \\ & = & 2M_A\prod_{B\neq A}\left[
\frac{1+\lambda^i_{A}\lambda^i_B}{1-\lambda^i_A\lambda^i_B}
\right]^{\balpha^A \cdot\balpha^B}
\end{eqnarray}
The expression for the determinants clearly diverges as the vevs approach 
the CMS i.e. $\lambda^i_A=\lambda^i_B$. 
This is not unexpected. On the CMS the instanton moduli space enlarges 
and the quadratic fluctuation operators gain extra zero eigenvalues. 
The singularities are the result of treating these modes in Gaussian 
approximation. To make progress we must treat these modes exactly. 
We then expect the instanton calculation to yield zero when the vevs lie 
on the CMS, for the instanton wll have fermionic zero modes which cannot be 
saturated in the path integral.    

\subsection*{Instanton Soft Modes}

Close to the CMS, the modes that become zero modes are soft; that is the 
eigenvalues of the quadratic fluctuation operators are small. 
The existence of these modes is reminiscent of 
the more familiar situation of four dimensional instantons where a self-dual 
field strength ceases to satisfy the equations of motion when an adjoint 
scalar has a non-zero vev and the one-instanton moduli space is lifted 
leaving just the singular point at the origin. 
However the self dual configurations retain their 
importance in the semi-classical expansion. The correct technique 
for dealing with such modes is known as the constrained instanton 
\cite{aff} (For a detailed account applied to four dimensional 
${\cal N}=2$ theories see also \cite{mo1}). At short distances the equations 
of motion are solved perturbatively in $g^2\rho^2v^2$ where $\rho$ is 
the scale size of the instanton. This allows all self-dual configurations 
to be treated exactly in the semi-classical expansion. 
The action of these configurations gains a $\rho$ dependence 
ensuring the contribution 
of the larger instantons to the path integral are suitably suppressed.

The three dimensional situation is analagous. The vevs of 
$\hat{\phi}^i$ lift certain solutions to the equations of motion which 
still remain important in the semi-classical expansion near the CMS
\footnote{ This correspondence is emphasised further if we trace the 
three dimensional theory back to its ${\cal N}=2$ four dimensional roots, 
combining $A_\mu$ and $\Phi$ to construct a self-dual gauge field. The two 
independant degrees of freedom in $\hat{\phi}^i$ create the complex 
scalar field and the generic self-dual field strength no longer satisfies the 
equations of motion when this scalar has a vev.}. However, the details 
of the lifting of the instanton moduli spaces are more complicated. 
A zero vev for all 
$\hat{\phi}^i$ corresponds to the intersection of all the CMS and the moduli 
space of solutions is given by the single Higgs results of Weinberg 
\cite{wein}. Turning on vevs for for $\hat{\phi}^i$ generically means a 
departure from the CMS and the moduli space is lifted, although a non-trivial 
submanifold may remain. Moreover, by varying the vevs in special directions 
along the CMS intermediate situations are possible with submanifolds of 
exact solutions of varying dimensions. 

Employing the constrained instanton, we relax the conditions on the 
configurations about which we perform the semi-classical expansion. 
Rather than insisting configurations are a minimum of the action, in 
the short distance regime, $x\ll 1/M_A$, we solve the equations of 
motion perturbatively in an appropriate parameter, generically 
$e^2r^2\|\hat{\bf v}^i\|^2/\|{\bf v}^i\|^2$ 
where $r$ dentotes all radial parameters on the 
largest instanton moduli space. In the language of Weinberg \cite{wein}, 
for large $r$, they are the collective coordinates 
obtained by separating two ``fundamental'' instantons. 

The approach of solving the equations perturbatively also applies to the 
auxillary fields which we have so far neglected. There exist three auxilary 
fields, $F^i$, one for each $N=1$ scalar multiplet. There are no 
auxillary fields from gauge multiplets in three dimensions. 
In the $N=4$ theory $F^i$ satisfy the equation of motion
\begin{equation}\label{aux}
F^i=-i\epsilon^{ijk}\phi^j\phi^k
\end{equation}
Writing $\hat{F}^i=F^i-\lambda^i_A\lambda^j_AF^j$, the defining equations 
for the constrained instanton at short distance are given by, 
\begin{eqnarray}
{\cal D}_\mu\Phi & = & B_\mu \label{bog1} \\
\gamma_\mu {\cal D}_\mu\chi^m-[\Phi ,\chi^m] & = & 0 \\
(P_-)_{mn}\chi^n & = & \chi^m \\
{\cal D}_\mu {\cal D}_\mu\hat{\phi}^i-[\Phi ,[\Phi ,\hat{\phi}^i]] & = & 
-\eta^i_{mn}\chi^m\chi^n\label{bg} \\ 
\hat{F}^i & = & i(\lambda^i_A\lambda^j_A\epsilon^{jkl} - 
\epsilon^{ikl})
\phi^k\phi^l \\ 
\lambda^i_A F^i & = & 0 \label{end}
\end{eqnarray}

The bosonic 
moduli space of such solutions is determined  solely by solutions to 
(\ref{bog1}). In the topological sector ${\bf g}=\balpha^A$, the vev 
$\lambda^i_A{\bf v}^i$ picks auxillary simple roots, $\bgamma^a$ s.t. 
$\balpha^A=\sum_a m_a\bgamma^a$. The moduli space of solutions is 
of dimension $4\sum_am_a$ \cite{wein}. Notice that the roots 
$\bgamma^a$ differ from sector to sector and need not coincide with 
the $\bbeta^a$ defined earlier. 
This means the relative dimensions of the moduli spaces of (\ref{bog1}) 
in different sectors do not follow the 
simple pattern of the single Higgs model. Moreover, in varying the 
${\bf v}^i$ it is possible for $\lambda^i_A{\bf v}^i$ to cross the wall 
of a Weyl chamber without the associated non-maximal symmetry breaking 
of the single Higgs model and thus the moduli space  
may change discontinuously. However, after integration over these 
manifolds, the final instanton calculation will be smooth.

The general moduli space decomposes into the form
\begin{equation}\label{moduli}
{\cal M}={\bf R}^3\times\frac{{\bf R}\times\tilde{\cal M}_d}{Z}
\end{equation}
$\tilde{\cal M}_d$ are complete HyperK\"ahler manifolds with coordinates 
$X^a$ and metric $\tilde{g}_{ab}$. For ${{\bf g}=\balpha^A}$, 
$\tilde{\cal M}_d$ are the Lee-Weinberg-Yi spaces \cite{lwy}. 
$\tilde{\cal M}_d$ has dimension $4(d-1)$ where 
$d$ is the height of the root $\balpha^A$ as measured by $\bgamma^a$ i.e. 
$d=\sum_am_a$. 
In the standard notation, 
$\tilde{\cal M}_d=\tilde{\cal M}_{(1,1,..,1)}$ where there are 
$d$ 1's in the string. For $\balpha^A$ simple with respect to 
$\lambda^i_A{\bf v}^i$ ($d=1$), $\tilde{\cal M}_1$ 
is taken to be a single point.

The ${\bf R}^3$ factor in (\ref{moduli}) corresponds to space-time 
translations of the instanton while the ${\bf R}$ factor is generated 
by global $U(1)$ gauge transformations 
\begin{equation}
{\bf Q}\cdot{\bf H}=\frac{\sum_a(\lambda^i_A{\bf v}^i\cdot\bgamma^a)\,
\bomega^a\cdot{\bf H}}{\sum_b\lambda^i_A{\bf v}^i\cdot\bgamma^b}
\end{equation}
where $\bomega^a$ are the fundamental weights defined by 
$\bgamma^a\cdot\bomega^b=\frac{1}{2}\delta^{ab}$. 
When the ratios of the ${\lambda^i_A{\bf v}^i\cdot\bgamma^a}$ are rational 
the ${\bf R}$ factor collapses to $S^1$ and the $Z$ to the cyclic subgroup 
$Z_d$. 

The remaining $n-2$ $U(1)$ gauge 
transformations generated by elements of the CSA orthogonal to 
$\balpha^A$ result in up to $n-2$ $U(1)$ isometries 
of $\tilde{\cal M}$. We denote as ${\bf K}^a$ the Killing vector of 
${\cal M}$ generated by ${\bf H}$. If a particular element of the CSA acts 
trivially on the configuration, the corresponding Killing vector is taken  
to be zero. 

Configurations satisfying (\ref{bog1}-\ref{end}) raise the action 
above the Bogomol'nyi bound (\ref{action}). Moreover, this action will 
have dependance on the collective coordinates of $\tilde{\cal M}_d$; 
we can consider the action as defining a potential on $\tilde{\cal M}_d$. 
The bosonic contribution to this potential is  
\begin{eqnarray}\label{action2}
\Delta S_{\rm bose} & = & \frac{2\pi}{e^2}\int d^3x\ 
\frac{1}{2}{\cal D}_\mu\hat{\phi}^i{\cal D}_\mu \hat{\phi}^i + \frac{1}{2}
F^iF^i \nonumber \\
 & = & \frac{2\pi}{e^2}\int d^3x\ \frac{1}{2}{\cal D}_\mu
\hat{\phi}^i{\cal D}_\mu \hat{\phi}^i - \frac{1}{2}[\Phi ,\hat{\phi}^i]^2
\end{eqnarray}
We will now show this potential is related to the $U(1)$ Killing vectors 
on the moduli space ${\cal M}$. Firstly 
consider a configuration of $A_\mu$ and $\Phi$ satisfying (\ref{bog1}) 
and thus corresponding to a point ${\cal M}$. 
Act on these fields with a gauge transformation parameteretrised  by 
$\hat{\phi}^i$
\begin{equation}\label{gt}
\delta^iA_\mu = {\cal D}_\mu\hat{\phi}^i\ \ \ \ ;\ \ \ \ \delta^i\Phi
 = -i[\Phi , \hat{\phi}^i]
\end{equation} 
Such transformations clearly satisfy the linearised version 
of the Bogomol'nyi equation (\ref{bog1}). A global (large) gauge 
transformation is generated by $\hat{\bf v}^i\cdot{\bf H}$, 
the vev of $\hat{\phi}^i$, 
while the spatially dependant part is a small gauge transformation. 
However, in order to be true zero modes of the configuration, the 
transformations must also satisfy a gauge condition, which for the purely 
bosonic theory is Gauss' law
\begin{equation}\label{gauge}
{\cal D}_\mu\delta A_\mu-i[\Phi, \delta\Phi] =0
\end{equation}
We see that, up to fermi bilinears, this equation is indeed satisfied by 
virtue of equation (\ref{bg}). Thus the gauge transformations above 
are indeed true zero modes. Moreover, gauge transformations move 
the configuration infinitesimally along the isometries of ${\cal M}$. 
By construction $\balpha^A\cdot\hat{\bf v}^i=0$ 
and gauge transformations (\ref{gt}) correspond to elements of the CSA 
responsible for generating the Killing vectors on $\tilde{\cal M}_d$ 
together with a vev dependant shift along the ${\bf R}$ factor. 
Thus, the bosonic part of the action can be rewritten as
\begin{equation}\label{bosact}
\Delta S_{\rm bose}= \frac{4\pi^2}{M_Ae^2}\|{\bf Q}\cdot\hat{\bf v}^i\| + 
\frac{1}{2}\tilde{g}_{ab}(\hat{\bf v}^i\cdot{\bf K}^a)
(\hat{\bf v}^i\cdot{\bf K}^b)
\end{equation}
Equations (\ref{bog1}-\ref{end}) are covariant under supersymmetry 
transformations parametrised by $(P_+)_{mn}\xi^n$ ensuring that the total 
action of these configurations inherits a supersymmetry acting on the 
collective coordinates. Indeed, we may always replace the sum over  
Killing vectors in the second term of (\ref{bosact}) by a single 
Killing vector, $\tilde{g}_{ab}(\hat{\bf v}^i\cdot{\bf K}^a)
(\hat{\bf v}^i\cdot{\bf K}^b)=\tilde{g}_{ab}({\bf V}\cdot{\bf K}^a)
({\bf V}\cdot{\bf K}^b)$ which is the form dictated by supersymmetry 
\cite{ag}. The fermionic part of the action is 
simply the supersymmetric completion of (\ref{bosact}). 
\begin{equation}\label{feract}
\Delta S_{\rm fermi}=\frac{i}{2}\nabla_a({\bf V}\cdot
{\bf K}_b)\psi^a\psi^b
\end{equation}
where $\nabla_a$ is the covariant derivative on ${\cal M}$ with respect 
to the Levi-Civita connection and $\psi^a$ are fermionic collective 
coordinates. 

We turn finally to the measure. As neither the metric nor the integrand 
depend upon the coordinates associated with the $U(1)$ isometries, the 
bosonic and fermionic measure for the ${\bf R}^3\times{\bf R}$ 
factor of the moduli space are given (after taking into account the discrete 
group $Z$) by (\ref{bmeasure}) and (\ref{fmeasure}) respectively. This 
leaves us with the integrations over the multi-cover of the 
Lee-Weinberg-Yi space, $\tilde{\cal M}_d$. Although expressions for the 
corresponding zero modes are not known, the Jacobians depend on 
the metric, $\tilde{g}_{ab}$ only,
\begin{eqnarray}\label{measures}
\int\  d\tilde{\mu}_B & = & \int\ \prod_{a=1}^{4(d-1)} dX^a
\frac{\sqrt{{\rm det}\tilde{g}}}{(2\pi )^{2(d-1)}}
 \nonumber \\
\int\ d\tilde{\mu}_F & = & \int \ \prod_{a=1}^{4(d-1)} d\psi^a ({\rm det}
\tilde{g})^{-1/2}
\end{eqnarray}
Note that the metric dependance in the bosonic and fermionic measure 
cancels.

\subsection*{Instanton Calculation}

Having analysed the various fluctuations around the background of the 
instanton, it is now possible to put all the pieces together. 

In each topological sector, defined by $\balpha^A$, we 
must calculate the height, $d$, of $\balpha^A$ with respect to 
$\bgamma^a$. If $\lambda^i_A{\bf v}^i\cdot\balpha^B=0$ for some root 
$\balpha^B$, the vev lies on the wall of 
a Weyl chamber it does not define a unique set of $\bgamma^a$ and  
the height of $\balpha^A$ and hence the moduli space $\tilde{\cal M}_d$ 
is ambiguous. We will comment on this case at the end of this section. 
For now we assume the vev $\lambda^i_A{\bf v}^i$ 
lies strictly within a Weyl chamber. The integration over soft modes 
is then given by
\begin{equation}\label{soft}
L_d({\bf v}^i) = (2\pi)^{2(1-d)}\int_{\tilde{\cal M}_d}\ dX^ad\psi^a
\ \exp \left( -\Delta S_{\rm bose}-\Delta S_{\rm fermi}\right)
\end{equation} 
For $d=1$ we set $L_1=1$. 

To avoid overcounting, 
we must divide by the Gaussian approximation 
for these modes which we have already taken into account 
when integrating over non-zero modes (\ref{R}). The recipe for 
this is to transform to polar coordinates for $\tilde{\cal M}_d$ 
such that in the vicinity of $r=0$, the metric is of the form $\tilde{g}_{ab}
=\tilde{g}_{ab}^{\rm\ flat}\left(1+O(r^2)\right)$ where $r$ denotes 
all $d-1$ radial coordinates on the space. The Gaussian approximation 
requires a truncation of this metric to $\tilde{g}_{ab}^{\rm\ flat}$ 
with the corresponding truncation to the bosonic potential and fermionic 
potential. We will denote this integral as $G_d$, again with 
$G_1=1$. Note that the first term in (\ref{bosact}) is independant of 
collective coordinates and will be cancelled after division by the 
Gaussian approximation. 

For $d=2$ the relevant manifold is Taub-NUT space \cite{gl,lwy1}. In 
appendix B we calculate $L_2$ and its Gaussian appproximation.

With all zero mode fields now confined to an $SU(2)$ subgroup the 
remainder of the instanton calculation for the four fermi vertex now proceeds 
as in \cite{dkmtv}, with the resulting correlator related to the Riemann 
tensor of the monopole metric. We choose instead to calculate instanton 
contributions to the scalar propagator which will provide direct information 
about the inverse metric. To saturate the fermionic zero modes of the 
instanton, the scalars must themselves pick up fermi bilinears. Acting on 
$\phi^i$ with a finite supersymmetry transformation, 
$\exp(-\xi^mQ^m)$, yields 
\begin{equation}
\phi^i\rightarrow\phi^i-B_\mu\eta^i_{mn}\xi^m\gamma_\mu\xi^n.
\end{equation}
For $\xi^m$ an eigenvector of $P_-$ only $\hat{\phi}^i$ pick up these 
bilinears. $\Phi$ remains unchanged by the supersymmetry tranformations. 
The contributions from the $\balpha^A$ instanton are thus
\begin{eqnarray}\label{prop}
\langle \Phi^a\Phi^b\rangle_A & = & 0 \nonumber \\
\langle \Phi^a\hat{\phi}^{ib}\rangle_A & = & 0 
\end{eqnarray}
and
\begin{equation}\label{answer}
\langle \hat{\phi}^{ia}\hat{\phi}^{jb}\rangle_A = (\bbeta^a\cdot\balpha^A)
(\bbeta^b\cdot\balpha^A)\,\frac{L_d({\bf v}^i)}{G_d({\bf v}^i)}
\,\prod_{B\neq A} 
\left[\frac{1+\lambda^i_A\lambda^i_B}{1-\lambda^i_A\lambda^i_B}
\right]^{\balpha^A\cdot\balpha^B} 
\langle\hat{\phi}^i\hat{\phi}^j\rangle_{\lambda^i{\bf v}^i.\small{\balpha^A}}
\end{equation}
where $\langle\hat{\phi}^i\hat{\phi}^j\rangle_{\lambda^i{\bf v}^i\cdot
\small{\balpha^A}}$ is the scalar propagator in an $SU(2)$ gauge theory 
with vev $\lambda{\bf v}^i.\balpha^A$. It was shown explicitly in \cite{dkmtv} 
that this scalar propagator reproduces the leading order exponential 
corrections to the inverse Atiyah-Hitchin metric.  

Close to the CMS, $L_d/G_d$ is small and cancels the singularities 
in the product factor. Far from the CMS, $L_d/G_d$ is exponentially 
close to unity. Furthermore, in this regime these exponential 
deviations from unity are of the same magnitude as other effects that 
we have neglected such as two-loop perturbation theory around the background 
of the instanton and instanton-anti-instanton pairs. 

Finally we turn the the situation where $\lambda^i_A{\bf v}^i\cdot\balpha^B
=0$ for some root $\balpha^B$ i.e. $\lambda^i_A{\bf v}^i$ lies on 
the wall of a Weyl chamber. In this case the moduli space ${\cal M}_d$ is 
not well defined. However this occurs in a regime far from the CMS and thus 
corrections from the constrained instanton are not important here. 
Nevertheless, it is interesting to note that $L_d/G_d$ does indeed 
remain smooth as $\lambda^i_A{\bf v}^i$ crosses the wall of the Weyl chamber. 

\section{Monopole Moduli Spaces}

We now translate the results of the previous section into the metric on 
the moduli space of $n$ BPS $SU(2)$ monopoles. This $4n$ dimensional 
space has the form
\begin{equation}\label{mspace}
{\cal M}_n=R^3\times\frac{S^1\times\tilde{\cal M}_n}{Z_n}.
\end{equation}
$R^3$ corresponds to (Euclidean) space-time translations of the centre 
of mass and $S^1$ to global $U(1)$ gauge transformations. $\tilde{\cal M}_n$ 
is the relative n-monopole moduli space; it has dimension 
$4(n-1)$, is complete and hyperK$\ddot{\rm a}$hler.  

The perturbative sector of the three dimensional $SU(n)$ gauge theory 
reproduces the asymptotic metric on $\tilde{\cal M}_n$ \cite{ch}. In 
this regime monopoles interact pairwise via velocity dependant $U(1)$ 
electric, magnetic and scalar forces and the metric takes a simple form 
discovered by Gibbons and Manton \cite{gm}. 
\begin{equation}\label{gmmetric}
ds^2=M_{ij}d\vec{x}_i\cdot d\vec{x}_j+M^{-1}_{ij}\left( d\theta_i
+\sum_{k}\vec{W}_{ik}\cdot d\vec{x}_k\right)\left( d\theta_j +
\sum_l\vec{W}_{jl}\cdot d\vec{x}_l\right)
\end{equation}
where
\begin{eqnarray}\label{gmdefs}
M_{ii}=1-\sum_{j\neq i}\frac{1}{r_{ij}}\ \ \ & ; &\ \ \ M_{ij}=\frac{1}{r_{ij}}
\ \ \ (i\neq j) \nonumber \\ \vec{W}_{ii}=-\sum_{j\neq i}\vec{w}_{ij} 
\ \ \ & ; &\ \ \ \vec{W}_{ij}=\vec{w}_{ij}\ \ \ (i\neq j)
\end{eqnarray}
with the Dirac potential, $\vec{w}_{ij}$, defined be ${\nabla}_i\times
\vec{w}_{ij}=\nabla_i (1/r_{ij})$. The $\vec{x}_i$ are the positions of 
the well-separated monopoles. This is the metric on ${\cal M}_n$; that 
is it includes the motion of the centre of mass and centre of charge 
of the monopole configuration. This corresponds to the three dimensional 
$U(n)$ gauge theory. In order to compare with the $SU(n)$ results above 
we must freeze the centres of mass and charge from this metric. 
For well seperated monopoles, the $4\times (n-1)$ coordinates on this 
space are a basis chosen from the 
$4\times\frac{1}{2}n(n-1)$ relative seperations and relative charges,
\begin{equation}
\vec{r}_{ij}=\vec{x}_i-\vec{x}_j\ \ \ \ ;\ \ \ \ \psi_{ij}=\theta_i-\theta_j.
\end{equation} 
An explicit hyperK$\ddot{\rm a}$hler quotient of the Gibbons-Manton metric 
yields a messy result, essentially because the metric is symmetric in all 
$\frac{1}{2}n(n-1)$ of the relative coordinates but is expressed in only an
$(n-1)$ dimensional subset of these. 
To retain manifest permutation symmetry of the relative-monopole metric we 
choose to write it as a metric on a larger 
$4\times\frac{1}{2}n(n-1)$ dimensional manifold, the 
pull-back of which yields the required quotient of the Gibbons-Manton metric. 
We take the relative distances, $r^A$, relative Euler angles, 
$\theta^A$ and $\phi^A$ and relative charges, $\psi^A$, $A=1,..,
\frac{1}{2}n(n-1)$ and write the metric in the form
\begin{equation}\label{metric}
ds^2=\frac{1}{2}\sum_A f^2(r^A)dr^Adr^A+a^2(r^A)
(\sigma_1^A)^2+b^2(r^A)(\sigma_2^A)^2+c^2(r^A;r^B)(\sigma_3^A)^2
\end{equation}
where all summations have been kept explicit. $f,a,b$ and $c$ take the form
\begin{eqnarray}\label{fabc}
 && \qquad f(r^A)=-\left( \frac{2}{n}-2M_A\right)^{\frac{1}{2}} \nonumber \\
 && a(r^A)=b(r^A)=r^A\left( \frac{2}{n}-2M_A\right)^{\frac{1}{2}} \\
 && \ \ \ c(r^A;r^B)=\left( \frac{1}{2n}-\frac{1}{2}(M^{-1})_A
\right)^{-\frac{1}{2}} \nonumber
\end{eqnarray}
where $M_A=M_{ij}$ for $A$ labelling the seperation $(ij)$. The 
one-forms $\sigma_i^A$ are defined as
\begin{eqnarray}
\sigma_1^A & = & -\sin\psi^Ad\theta^A + \cos\psi^A\sin\theta^Ad\phi^A 
\nonumber \\
\sigma_2^A & = & \cos\psi^Ad\theta^A + \sin\psi^A\sin\theta^Ad\phi^A \\
\sigma_3^A & = & d\psi^A+\frac{2}{n-1}\sum_B\,\Omega^{AB}
\cos\theta^B d\phi^B \nonumber
\end{eqnarray}
where $\Omega^{AB}$ is non-zero only if the seperations $A$ and $B$ 
have a monopole in common. More precisely, if the relative 
Cartesian seperation vector $\vec{r}^A$ goes from the $i^{th}$ to the 
$j^{th}$ monopole and $\vec{r}^B$ from the $k^{th}$ to the $l^{th}$,
\begin{equation}\label{Omega}
\Omega^{AB}=\frac{1}{2}(\delta^{ik}+\delta^{jl}-\delta^{il}-\delta^{jk})
\end{equation}

To recover the relative-monopole Gibbons-Manton metric, we must 
first pick a linearly 
independant set of $n-1$ seperations (3-vector and charge). Labelling the 
monopoles in some arbitrary manner, we choose the seperations between the 
$i^{\rm th}$ monopole and the ${(i+1)^{\rm th}}$. Labelling this set of 
linearly independant coordinates with a subscript $a$, ${(\vec{r}^{a}, 
\psi^a)}$, the metric on $\tilde{\cal M}_n$ is obtained by pulling back 
with the map, $f^A_a$, relating $\vec{r}^A$ and $\vec{r}^a$. \footnote
{Pulling back in polar coordinates is extremely messy. The simplest way 
to see this result is to first add terms corresponding to the 
centre of mass and charge, change to Cartesian coordinates and pull back 
to recover the orignal Gibbons-Manton metric.}
\begin{equation}\label{fmap}
\vec{r}^{A}=f^A_a\vec{r}^{a}
\end{equation}
Pulling back the flat metric gives the inverse Cartan matrix that was 
the classical metric of (\ref{claslow}),
\begin{equation}
\sum_A f^A_af^A_b=\frac{n}{2}(K^{-1})_{ab}.
\end{equation}
At this stage we need to introduce a dictionary between objects in 
three dimensions and the above coordinates on the monopole moduli space. 
Firstly, we note that the choice of seperations between $n$ monopoles 
discussed above is mimicked 
by the root structure of $su(n)$. The map $f^A_a$ is the map between 
simple roots, $\bbeta^a$ and roots, $\balpha^A$. Defining the roots 
as $\balpha^A=({\bf e}^i-{\bf e}^j)/\sqrt{2}$ and 
$\balpha^B=({\bf e}^k-{\bf e}^l)/\sqrt{2}$, where ${\bf e}^i$ are $n$ 
orthonormal vectors, then ${\Omega^{AB}=\balpha^A\cdot\balpha^B}$. 
We must identify the 
three scalar vevs along the $\balpha^A$ direction of root space 
with the vector distance between the $i^{\rm th}$ and $j^{\rm th}$ 
monopole \footnote{ To retain agreement between 
(\ref{claslow}) and (\ref{metric}), the metric should be premultiplied 
by $e^2/16n\pi^3$. This will not be important in what follows.}. 
\begin{eqnarray}\label{dict}
\vec{r}^{\,a}=r^{ai} & = & \frac{8\pi^2}{e^2}{\bf v}^i\cdot\bbeta^a 
\nonumber \\  
\psi^a & = & \frac{1}{2}{\bf \sigma}\cdot\bbeta^a 
\end{eqnarray}
Other expressions from the instanton calculation also have a simple 
geometrical meaning. In the topological sector, ${\bf g}=\balpha^A$, 
we have $\lambda^i_A=r^{Ai}/r^A$ and 
\begin{eqnarray}\label{dict2}
{\bf v}^i\cdot{\bf v}^i & = & \left(\frac{e^2}{8\pi^2}\right)^2\frac{2}{n}
\sum_B(r^B)^2 \\
\hat{\bf v}^i\cdot\hat{\bf v}^i & = & \left(\frac{e^2}{8\pi^2}\right)^2
\frac{2}{n}\sum_B (r^B_\perp )^2 \nonumber
\end{eqnarray}
where $r^{Bi}_\perp$ is the component of $r^{Bi}$ perpendicular 
to $r^{Ai}$.

The semi-classical approximation translates to the requirement 
that the monopoles are well seperated; we will find the leading 
order corrections to the Gibbons-Manton metric. 
Geometrically, the curves of marginal stability correspond to 
three or more monopoles becoming colinear. In this regime 
the corrections from the constrained instanton (\ref{soft}) 
will have most impact. The other point of interest is when the vev 
$\lambda^i_A{\bf v}^i$ lies on the wall of a Weyl chamber. This 
corrsponds to $r^{Ai}r^{Bi}=0$ for some seperation $B$. The 
form of the corrections to the metric will be different on each  
side of this situation but will meet smoothly on the wall itself.

We are now in a position to recast the 
instanton contribution to the scalar propagator (\ref{answer}) as the 
exponential corrections to the $n$-monopole metric. Corrections to 
the scalar propagator equate to corrections to the inverse metric
\begin{equation}
\delta g^{ai\, bj}=\sum_A(\balpha^A\cdot\bbeta^a)(\balpha^A\cdot\bbeta^b)
\langle\phi^i\phi^j\rangle_A
\end{equation}
To leading order in $1/r$, the inverse of this is
\begin{equation}
\delta g_{ai\, bj}=-(K^{-1})_{ac}(K^{-1})_{db}\sum_A(\balpha^A\cdot
\bbeta^c)(\balpha^A\cdot\bbeta^d)\langle\phi^i\phi^j\rangle^{-1}_A 
\end{equation}
which can be written simply as the pull-back of the diagonal metric with 
entries $\langle\phi^i\phi^j\rangle^{-1}_A$. The 
first exponential corrections to the functions $f,a,b$ and $c$ are thus 
given by
\begin{eqnarray}\label{fabc2}
 && \qquad \qquad \qquad
f(r^A;{\vec r}^{B},\psi^B)=-\left(\frac{2}{n}-2M_A\right)^{\frac{1}{2}} 
\nonumber \\
 && a(r^A;{\vec r}^{B},\psi^B)=r^A\left(\frac{2}{n}-2M_A\right)^{\frac{1}{2}}
-4(r^A)^2e^{-r^A}\frac{L_d}{G_d}\,\prod_{B\neq A}\left(\frac{1+
\cos\Theta_{AB}}{1-\cos\Theta_{AB}}\right)^{\Omega^{AB}} \nonumber \\ 
 && b(r^A;{\vec r}^{B},\psi^B)=r^A\left(\frac{2}{n}-2M_A\right)^{\frac{1}{2}}
+4(r^A)^2e^{-r^A}
\frac{L_d}{G_d}\prod_{B\neq A}\left(\frac{1+\cos\Theta_{AB}}{1-
\cos\Theta_{AB}}\right)^{\Omega^{AB}} \\ 
 && \qquad \qquad \qquad 
c(r^A;{\vec r}^{B},\psi^B)=\left(\frac{1}{2n}-\frac{1}{2}(M^{-1})_A
\right)^{-\frac{1}{2}} \nonumber
\end{eqnarray}
where $\Theta_{AB}=\hat{r}^{iA}\hat{r}^{iB}$.

Some comments are probably in order. Firstly, the product factor in 
(\ref{fabc2}) corresponds to non-pairwise scattering of monopoles. The 
interaction between a pair of monopoles depends on the distances between 
the pair and all other monopoles. As one monopole is taken to infinity, 
the corresponding part of the product tends towards unity and the 
monopole decouples from the interaction as expected. As monopoles 
become colinear, the product factor in (\ref{fabc2}) becomes 
singular, but this singularity is cancelled by $L_d/G_d$ and 
in each case the overall correction to the metric is zero. 

The situation of 
three monopoles is depicted in the figure. 
For the interaction between the 
first and second monopoles, the moduli space for the constrained instanton 
junps discontinuously when $\lambda^i_{(12)}{\bf v}^i\cdot\balpha^{(12)}
=|\lambda^i{\bf v}^i|\cos(\pi /3)$. In the monopole picture. this 
corresponds to $\alpha_{23}=\pi /2$ or $\alpha_{13}=\pi /2$. Thus, 
if both $\alpha_{23}$ and $\alpha_{13}$ are acute, as in trinagle 
(a), the correction from the constrained instanton is $L_2/G_2$. 
If either is obtuse as in triangle (b) the correction is 
$L_1/G_1=1$.

\begin{figure}[htp]
\begin{center}
\leavevmode
\centerline{\hbox{\epsfxsize=12 cm \epsfbox{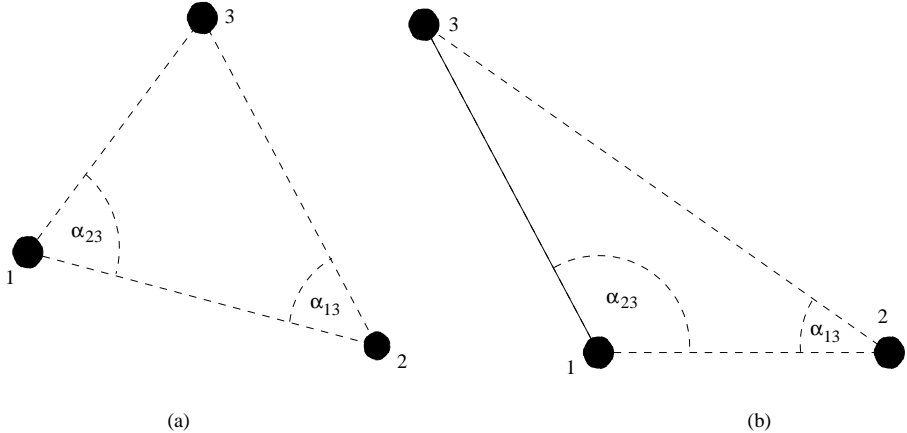}}}
\caption{For triangle (a), the height of the root cooresponding to the 
interaction between the first and second monopoles is 2. For triangle 
(b), the root is simple.}
\end{center}
\end{figure}

The product factor of equation (\ref{fabc2}) is
\begin{equation}\label{fig}
\left(\frac{1+\cos\alpha_{23}}{1-\cos\alpha_{23}}\right)^{\frac{1}{2}}
\left(\frac{1-\cos\alpha_{13}}{1+\cos\alpha_{13}}\right)^{-\frac{1}{2}}
\end{equation}
For triangle (a), $L_2/G_2$ can be read off from equation 
(\ref{l2g2}) of appendix B,
\begin{equation}\label{help} 
1-\left(1+\frac{1}{18}r_{12}\tan\alpha_{13}\tan\alpha_{23}\right)
\exp\left(-\frac{1}{18}r_{12}\tan\alpha_{13}\tan\alpha_{23}\right)
\end{equation}

As the third monopole is brought between the other two, the singularity 
in equation (\ref{fabc2}) is cancelled by (\ref{help}) and the 
interaction between the first and second vanishes.
As the third monopole is taken to infinity, $\alpha_{23}=\pi -\alpha_{13}$, 
the two factors in (\ref{fig}) cancel. The corrections 
from the constrained instanton (\ref{help}) retain some 
information about the position of the third monopole. However, the 
window in which such corrections are to be applied becomes vanishingly small 
and the third monopole decouples from the interaction as expected. 

Note that for a right angle triangle, equation (\ref{help}) is unity and thus 
we have a smooth transition between the acute and obtuse triangles. 
However, we once again emphasise that in this regime the exponential 
corrections in (\ref{help}) are subleading with respect to other 
corrections that we have not dealt with. 

The lack of exponential corrections to $f$ may appear odd to 
anyone familiar 
with the expansion of these functions for the case of two monopoles given 
in \cite{gm2} where $f$ also has exponential corrections 
of the same order as $a$ and $b$. These corrections certainly do not 
come from the instanton sector of the gauge theory as such terms come complete 
with ${\rm exp}(\pm i\sigma)$. The resolution of this matter is a problem 
common in dealing with exact 
results in SUSY gauge theories, namely that the vacuum moduli space 
is parametrised by coordinates of the low energy theory which have a 
complicated dependance on the coordinates defined in the original 
Lagrangian. Such behaviour is seen to arise in finite ${\cal N}=2$ theories 
in four dimensions \cite{dkm2}. In the present case, it means the first 
of the relations (\ref{dict}) must be corrected by powers of 
${\rm exp}(-r)$ in order to reproduce leading order exponential 
terms to $f(r^A)$. It is possible to acheive this while only affecting 
$a(r^A)$ and $b(r^A)$ at subleading order, corresponding to two loop 
perturbation theory about the background of the instanton.

It would be gratifying to compare (\ref{fabc2}) with known metrics on 
subspaces of the full moduli space. The first situation where 
such metrics are known is the case with monopoles co-linear 
and equidistant. For co-linear monopoles, there are only interactions between 
adjacent monopoles which are independant of the positions of the others 
and the corrections are just of the form $r\exp(-r)$. If we further impose 
equidistance, we are left with the leading order corrections of the 
Atiyah-Hithin metric in agreement with \cite{sutb}

The one other case in which exact metrics are known is 
for four monopoles. The metric on the one dimensional submanifold 
of tetrahedrally symmetric manifolds has been computed using Nahm data 
and is found to have the leading exponential corrections occuring at 
$\exp (-2r)$ \cite{bradsut}. From the three dimensional perspective, 
instanton contributions to this metric are of the form $\langle\Phi\Phi
\rangle$ and so vanish. Naively, it appears that the two pictures 
are in agreement, the $\exp (-2r)$ term corresponding to an 
instanton-anti-instanton pair. However, the same coordinate problems arise 
as with the $f(r)$ term in the Atiyah-Hitchin metric and the conclusion 
is that, while consistent, agreement between the two remains ambiguous.

\section{N=8, SU(n)}

In this final section, we turn our attention to the $N=8$ theory. 
Dimensionally reducing the 
${\cal N}=1$ ten dimensional theory to 3 dimensions, the field content 
of the $N=4$ theory is augmented by the addition of 4 scalars and 4 
Majorana fermions. While the alegbra has a $Spin(8)$ automorphism group, 
only a $Spin(7)$ R-symmetry is manifest in the Lagrangian description 
 with the vector transforming in a singlet, the scalars in ${\bf 7}$ and 
the fermions in ${\bf 8}$ \cite{s}. 

In four dimensions, 
a non-renormalisation theorem for the ${\cal N}=4$ theories prevents 
instanton corrections to eight fermi vertices \cite{ds,dkmsw}. This is 
no longer the case in three dimensions and eight fermi (or four 
derivative) vertices \cite{ds,pp,dkm}. The counting 
of zero modes proceeds as in section 2; ${\bf g}=k\balpha^A$ 
generically has $2k$ fermi zero modes while ${\bf g}$ 
not aligned with a root has none. Instantons 
in the latter sector again break all the supersymmetries and so fail 
to contribute. However, the addition of adjoint massless 
matter multiplets enhancing $N=4$ supersymmetry to $N=8$ 
allows for the lifting of zero modes not protected by supersymmetry and 
sectors labelled by ${\bf g}=k\balpha^A$ contribute for all $k$ and all 
$\balpha^A$. 

The instanton calculation now proceeds identically to 
the $SU(2)$ case. The reader is referred to \cite{dkm} for details. 
The lifing of the zero modes is such that integration over them 
yields the volume contribution to the Euler character of the 
relative instanton moduli space, which generically for ${\bf g}=k\balpha^A$ 
is the $\tilde{\cal M}_k$ of (\ref{mspace}). This can be identified with 
the Euler character of $\tilde{\cal M}_k$ only up to boundary terms. 
For the case $k=2$, the metric is known explicity and the 
boundary terms vanish \cite{gh}. For higher charges, the Gibbons-Manton 
metric corresponding to well-separated monopoles has the required 
asymptotic flatness for the boundary term to vanish, 
but there may be contributions from ``clustering regions'' at the 
boundary where at least one pair of monopoles remain close. As 
in \cite{dkm}, we assume that this is not the case and that 
the integral over zero modes does indeed yield the Euler character. 

The Euler characters of the k-monopole $SU(2)$ relative monopole 
moduli spaces are determined to be \cite{ss}, 
\begin{equation}
\chi\left(\tilde{\cal M}_k\right) = k
\end{equation}

Importantly, just as in the $SU(2)$ case, 
the integrations over non-zero fluctuations about 
the background of the instanton, which were ultimately responsible 
for the non-pairwise interaction of monopoles in the $N=4$ case, 
now give unity. The singularities which occured on the CMS 
are no longer there. It is plausible that the extra supersymmetry 
takes care of the soft modes circumventing the need for the 
constrained instanton approach of section 2, although a proof of this 
would require knowledge of the Euler characters and boundary terms 
of moduli spaces for higher rank gauge groups. Let us see how this 
might occur.

Suppose we do not use the constrained instanton. Then on the CMS 
the instanton moduli space jumps discontinuously from 
$\tilde{\cal M}_k$ to $\tilde{\cal M}_{k(1,1,..,1)}$; that 
is from an $SU(2)$ to an $SU(d)$, $d\leq n$, moduli space. Unlike the 
$N=4$ case, the extra fermionic zero modes are lifted 
and the resulting integration becomes the Euler 
character of the enlarged moduli space (up to the complication 
of boundary terms). The instanton contribution 
to the correlation function is continuous over the CMS provided
\begin{equation}\label{euler2}
\chi\left(\tilde{\cal M}_{k(1,1,..,1)}\right) = \,\chi\left(
\tilde{\cal M}_{k}\right)=k
\end{equation}
This scenario remains more or less the same if we do use the machinery 
of the constrained instanton. The integration over soft modes and zero 
modes now yields the volume contribution to the G-index generalisation 
of the Gauss-Bonnet integral \cite{ag} and again, up to boundary terms, 
 continuity of the instanton calculation requires the Euler 
characters to be given by (\ref{euler2}).

There are also other instanton contributions which occur only on the CMS; 
namely those with charges not proportional to a root. Recall, 
for the generic situation the actions of such instantons are lifted 
above the Bogomol'nyi bound. On the CMS, the 
contributions from such solutions are again proportional to the 
Euler character of the moduli space. For continuity we require this to 
vanish. It is interesting to note that these spaces are expected 
to contain no square-integrable harmonic forms. While not a 
direct prediction of S-duality \cite{dfhk}, were these forms to exist, 
states in the four dimensional ${\cal N}=4$ theory would appear at no point 
in the moduli space except on the curves of 
marginal stabiltiy, where they are at their most vulnerable for decay. 
 
Finally we turn to a rather different application of three dimensional 
gauge theories. Polchinski and Pouliot \cite{pp} relate the dynamics 
of $N=8$ $SU(2)$ theory to the scattering of two membranes in Matrix theory. 
Instanton processes of charge $k$ correspond to scattering with momentum 
exchange in the eleventh direction, giving an important test of the 
eleven dimensional Lorentz invariance of Matrix theory. The $SU(n)$ 
theory considered here is of course related 
to the scattering of $n$ membranes. 
The transverse distances and interactions between branes follow the 
same pattern as the monopoles in the $N=4$ theory. To model moving 
membranes, the vevs are allowed time dependence and the four 
time derivative vertex of the low-energy action becomes a quartic 
velocity term in the low velocity effective action for interacting 
membranes. Usually such actions only make sense up to terms quadratic 
in the velocity, but for purely gravitational systems and systems 
with constant charge to mass ratio, the backreaction from the fields 
enters at the $5^{th}$ order in the velocities and the expansion may 
continue to quartic terms (see pages 165 and 337 of \cite{ll}). 
Indeed, in the present context 
the moduli space of membrane solutions in eleven dimensional supergravity 
is flat and the quadratic terms vanish. 

The cancellation of the non-zero mode fluctuations ensures that the 
four derivative term is just the sum over pairs of membranes. Unlike 
the monopole case, the longitudinal scattering of membranes in Matrix theory 
occurs pairwise. It would be interesting to see if this behaviour is 
reproduced in supergravity.

\centerline{*******************}

We are grateful to  H. Braden, G. Gibbons, V. Khoze, R. Mainwaring, 
N. Manton, S. Vandoren and especially N. Dorey and 
T. Hollowood for useful discussions. Both authors are supported 
by PPARC studentships.

\section*{Appendix A: Instanton Zero Modes}

In this appendix we calculate number of zero modes of the Dirac equation 
in the background of an instanton with several Higgs fields. 
We follow Weinberg \cite{wein} closely, concentrating on points that 
differ from the original calculation. Firstly we define a set of $16\times 16$ 
6-dimensional gamma matrices (this unconventional representaion arises 
from using t'Hooft matrices, $\eta^i$, rather than Pauli matrices, $\sigma_i$)
\begin{equation}\label{6gamma}
\Gamma_{\mu =1,2,3}=-\gamma_\mu\otimes 1\otimes\sigma_2\ \ \ ;\ \ \ 
\Gamma_{i=4,5,6}=-i\otimes \eta^i\otimes\sigma_1
\end{equation} 
and $\Gamma_7=1_2\otimes 1\otimes\sigma_3$. Writing $\Delta_M$ 
as the 6-dimensional covariant derivative
\begin{equation}\label{GammaDelta}
\Gamma_M\Delta_M=\left(\begin{array}{cc} 0 & \Delta \\ -\Delta^{\dagger} & 0 
\end{array}\right)
\end{equation}
where $\Delta$ is defined in equation (\ref{dirac}) and the explicit matrix in 
(\ref{GammaDelta}) refers to the last of the three direct products. 
We now rewrite (\ref{Imu}) as
\begin{equation}\label{Imu2}
I(\mu^2)=-{\rm Tr}_-\left(\Gamma_7\frac{\mu^2}{-(\Gamma\cdot\Delta)^2+\mu^2}\right)
=-{\rm Tr}\left( P_-\Gamma_7\frac{\mu^2}{-(\Gamma\cdot\Delta)^2+\mu^2}
\right)
\end{equation}
Using 
\begin{equation}
P_-\lambda^i_{\bf g}\eta^i=\lambda^i\eta^iP_-\ \ \ \ ;\ \ \ \ 
P_-(\eta^i-\lambda^i_{\bf g}\lambda^j_{\bf g}
\eta^j) = (\eta^i-\lambda^i_{\bf g}\lambda^j_{\bf g}\eta^j)P_+
\end{equation}
and tracing liberally over $\eta$ and $\gamma$ matrices, we may 
rewritten as
\begin{eqnarray}\label{Imu3}
I(\mu^2) & = & \int d^3x\ {\rm tr} \ P_-\Gamma_7\Gamma_\mu{\cal D}_\mu\,
\langle x|(\Gamma\cdot\Delta +\mu)^{-1}|x\rangle \\ & & + 
\int d^3x\ {\rm tr} \ \frac{i}{2}P_-\Gamma_7\Gamma_i\,\langle x|\hat{\phi}^i
(\Gamma\cdot\Delta +\mu)^{-1}|x\rangle
\end{eqnarray}
where ${\rm tr}$ denotes the trace over group and (6-dimensional) spinor 
indices only. The second term differs from Weinberg's calculation 
and is due to the extra Higgs fields. We multiply on top and bottom 
by $-\Gamma .\Delta +\mu$, take spinor 
traces and move this term to the left hand side of the equation, to 
arrive at 
\begin{equation}\label{Imu4}
-{\rm Tr}\left(P_-\Gamma_7\frac{\mu^2+\hat{\phi}^i\hat{\phi}^i}
{-(\Gamma\cdot\Delta)^2+\mu^2}\right)=-\int d^3x\ {\rm tr}\left( P_-\Gamma_7
\Gamma_\mu{\partial}_\mu\,\langle x|(\Gamma\cdot\Delta +\mu)^{-1}
| x\rangle\right)
\end{equation}
At this point it is important to note that although we have traced 
over spinor indices extensively, the derivation of (\ref{Imu4}) 
does not rely on the trace over group or spatial indices, 
allowing us to divide by the numerator of the left hand side. Integration 
by parts then yields
\begin{eqnarray}\label{Imu5}
I(\mu^2) & = & \int_{\Sigma_\infty}d^2S^\mu\ {\rm tr}\left( P_-\Gamma_7
\frac{\mu^2}{\hat{\phi}^i\hat{\phi}^i+\mu^2}\Gamma_\mu\frac{1}{\Gamma\cdot
\Delta +\mu}
\right)\\ & & -\int d^3x\ {\rm tr}\left( P_-\Gamma_7\Gamma_\mu{\cal D}_\mu
\left(\frac{\mu^2}{\hat{\phi}^i\hat{\phi}^i+\mu^2}\right)\frac{1}{\Gamma\cdot
\Delta +\mu}\right)
\end{eqnarray}
The second term vanishes using ${\cal D}_\mu\hat{\phi}^i=0$. 
The first term is very similar to the corresponding expression in Weinberg. 
Again multiply top and bottom by $-\Gamma .\Delta +\mu$ and perform 
the trace over the spinor indices. The presence of $P_-$ in 
the numerator kills all but the $\Phi$ term in $\Delta$. Expanding.  
\begin{eqnarray}
\frac{1}{-(\Gamma\cdot\Delta )^2+\mu^2} & = & \frac{1}{-{\cal D}^2+\phi^i\phi^i
+\mu^2} \nonumber \\ & & + \frac{1}{-{\cal D}^2+\phi^i\phi^i+\mu^2}2\gamma_\mu 
B_\mu\left(\small{\begin{array}{cc} P_+ & 0 \\ 0 & P_- \end{array}}
\right)\frac{1}{-{\cal D}^2+\phi^i\phi^i+\mu^2} \nonumber \\ & & + ...
\end{eqnarray}
and keeping terms of $O(1/x^2)$, evaluation of this first term now 
proceeds as Weinberg. Tracing over group indices indeed yields 
equation (\ref{337}) as claimed.

\section*{Appendix B: Integration Over Taub-NUT}

If $\balpha^A$ is of height 2, say $\balpha^A=\bgamma^1+\bgamma^2$, 
the relevant moduli space $\tilde{\cal M}_{(1,1)}$ is Taub-NUT \cite{gl,lwy1}.
Defining the reduced mass, $\mu = (\lambda^i_A{\bf v}^i
\cdot\bgamma^1)(\lambda^j_A
{\bf v}^j\cdot\bgamma^2)/M_A$, the metric is given by
\begin{equation}\label{TN}
{\rm d}s^2=\frac{8\pi^2}{e^2}\left(\mu V(r){\rm d}{\bf r}^2+\frac{1}{4}\mu^{-1}
V(r)^{-1}({\rm d}\psi +\cos\theta{\rm d}\phi )^2\right)
\end{equation}
where $0\leq\psi\leq 4\pi$ and
\begin{equation}
V(r)=1+\frac{1}{2\mu r}
\end{equation}
The Killing vector $\partial_\psi$ is generated by $(\bgamma^1-\bgamma^2)
\cdot{\bf H}/3$. Thus the action of the constrained instanton configurations 
parametrised by this space is raised by
\begin{equation}
\frac{1}{2}\tilde{g}_{ab}(\hat{\bf v}^i\cdot{\bf K}^a)
(\hat{\bf v}^i\cdot{\bf K}^b)=\frac{2\pi^2}{3e^2}\hat{\bf v}^i
\cdot\hat{\bf v}^i\frac{r}{1+2\mu r}
\end{equation}
together with the $\|{\bf Q}\cdot\hat{\bf v}^i\|$ term of equation 
(\ref{bosact}). 
Note that this potential flattens out as $r\rightarrow\infty$. This is 
a generic feature of all potentials on Lee-Weinberg-Yi spaces generated 
by the $U(1)$ Killing vectors. Because the Lee-Weinberg-Yi spaces are 
non-compact, the integral over this potential alone will diverge. 
The integral is rendered finite by the corresponding integration 
over fermionic coordinates. In the present case there is only one Killing 
vector on the moduli space and the potential is already in the 
form $({\bf V}\cdot{\bf K}^a)({\bf V}\cdot{\bf K}_b)$ with 
${\bf V}\cdot{\bf V}=\hat{\bf v}^i\cdot\hat{\bf v}^i$. The 
fermionic potential is given by
\begin{equation}
\sum_i\frac{i}{2}\nabla_a({\bf V}\cdot{\bf K}_a)\psi^a\psi^b = 
\frac{2i\pi^2}{\sqrt{3}e^2}\frac{(\hat{\bf v}^i\cdot\hat{\bf v}^i)^{
\frac{1}{2}}}{(1+2\mu r)^2}\left(\cos\theta\,\psi^r\psi^\phi+
\psi^r\psi^\psi-r\sin\theta(1+2\mu r)\psi^\theta\psi^\phi\right)
\end{equation}
Using the measures of integration (\ref{measures}), the integration 
over fermionic coordinates brings down two factors 
of the fermionic potential, isolating the $\psi^r\psi^\psi\psi^\theta
\psi^\phi$ term and leaving us with the following expression for 
$\tilde{L}_2=L_2\exp(+4\pi^2\|{\bf Q}\cdot\hat{\bf v}^i\|/M_Ae^2)$,
\begin{eqnarray}
\tilde{L}_2 & = & \frac{\pi^2}{3e^2}\hat{\bf v}^i\cdot\hat{\bf v}^i\int\ 
{\rm d}r{\rm d}\phi{\rm d}\theta{\rm  d}\psi\ r(1+2\mu r)^{-3}\sin\theta\ 
\exp\left[-\frac{2\pi^2}{3e^2}\hat{\bf v}^i\cdot\hat{\bf v}^i
\frac{r}{1+2\mu r}\right] 
\nonumber \\ 
& = & \frac{12}{\hat{\bf v}^i\cdot\hat{\bf v}^i}\left[
1-\left(1+\frac{2\pi^2}{3e^2}\frac{\hat{\bf v}^i\cdot\hat{\bf v}^i}
{2\mu}\right)\exp-\left(\frac{2\pi^2}{3e^2}\frac{\hat{\bf v}^i\cdot
\hat{\bf v}^i}{2\mu}\right)\right]
\end{eqnarray}
Notice that for small $\hat{\bf v}^i\cdot\hat{\bf v}^i$, $L_2
\sim(\hat{\bf v}^i\cdot\hat{\bf v}^i)^2$.

In order not to integrate over these modes twice, we must divide 
by the Gaussian approximation to $L_2$. The coordinate system used 
in (\ref{TN}) has a coordinate singularity at the origin, $r=0$. 
In order to present a metric that is smooth at the origin, we 
transform to the coordinate, $R$, where $r=\frac{1}{2}\mu R^2$. 
In this basis, the metric is
\begin{eqnarray}\label{TN2}
{\rm d}s^2 = \frac{8\pi^2\mu}{e^2}\big[ (1+\mu^2R^2){\rm d}R^2 
& + & \frac{1}{4}R^2(1+\mu^2R^2)({\rm d}\theta^2+\sin^2\theta
{\rm d}\phi^2) \nonumber \\ 
& + & \frac{1}{4}R^2(1+\mu^2R^2)^{-1}({\rm d}\psi^2+\cos\theta
{\rm d}\phi^2)\big]
\end{eqnarray}
Note that to leading order in $R$, this is the flat metric 
on ${\bf R}^4$ in Euler angle coordinates. The recipe for the 
Gaussian approximation is to truncate the Taub-NUT metric 
(\ref{TN2}) to the flat metric and repeat the calculation 
above using this. The Gaussian approximation to the bosonic 
potential is thus
\begin{equation}
\frac{\pi^2\mu}{3e^2}\hat{\bf v}^i\cdot\hat{\bf v}^i\,R^2\,\exp
-\left(\frac{4\pi^2\|{\bf Q}\cdot\hat{\bf v}^i\|}{M_Ae^2}\right)
\end{equation}
and the fermionic counterpart
\begin{equation}
\frac{2i\pi^2\mu}{\sqrt{3}e^2}(\hat{\bf v}^i\cdot\hat{\bf v}^i)^{
\frac{1}{2}}\left(R\cos\theta\,\psi^R\psi^\phi 
+ R\psi^R\psi^\psi-\frac{1}{2}R^2\sin\theta\,\psi^\theta\psi^\phi\right)
\end{equation}
Once more performing the integrations, we deduce $G_2=12/(
\hat{\bf v}^i\cdot\hat{\bf v}^i)\exp(-4\pi^2\|{\bf Q}\cdot{\bf H}\|/M_Ae^2)$ 
and thus,
\begin{equation}\label{l2g2}
\frac{L_2}{G_2}=1-\left(  1+\frac{\pi^2(\hat{\bf v}^i\cdot\hat{\bf v}^i)}
{3e^2\mu}\right)\exp\left[-\frac{\pi^2(\hat{\bf v}^i\cdot\hat{\bf v}^i)}
{3e^2\mu}\right]
\end{equation}

\end{document}